\documentclass[]{PoS}
\usepackage{natbib}

\newcommand{\Porb}{P_{\mathrm{orb}}}
\newcommand{\Msun}{M_{\odot}}

\newcommand{\Mwd}{M_{\mathrm{WD}}}

\newcommand{\mnras}{MNRAS}
\newcommand{\apj}{ApJ}
\newcommand{\apjs}{ApJSS}
\newcommand{\aap}{A\&A}

\title{Empirical consequential angular momentum loss can solve long standing problems of CV evolution}

\ShortTitle{CV evolution and CAML}

\author{\speaker{Matthias R. Schreiber}\\
        Universidad de Valpara\'iso, Instituto de F\'isica y Astronom\'ia, Avenida Gran Breta\~na 1111, Valpara\'iso, Chile\\}

\author{Monica Zorotovic\\
Universidad de Valpara\'iso, Instituto de F\'isica y Astronom\'ia, Avenida Gran Breta\~na 1111, Valpara\'iso, Chile\\}
\author{Thomas P.\,G. Wijnen\\
Department of Astrophysics/IMAPP, Radboud University Nijmegen, PO box
9010, NL-6500 GL Nijmegen, The Netherlands\\}

\abstract{
The observed orbital period distribution of cataclysmic variables (CVs), the space density derived from observations, and the observed orbital period minimum are known to disagree with theoretical predictions since decades. More recently, the white dwarf (WD) masses in CVs have been found to significantly exceed those of single WDs, which is in contrast to theoretical expectations as well.
We here claim that all these problems are related and can be solved if CVs with low-mass white dwarfs are driven into dynamically unstable mass transfer due to consequential angular momentum loss (CAML).
Indeed, assuming CAML 
increases as a function of decreasing white dwarf mass can bring into agreement the predictions of binary population models and the observed properties of the CV population. We speculate that a common envelope like evolution of CVs with low-mass WDs following a nova eruption might be the physical process behind our empirical prescription of CAML.  
}

\FullConference{The Golden Age of Cataclysmic Variables and Related Objects - III, Golden2015 \\
		7-12 September 2015\\
		Palermo, Italy}

\begin{document}

\section{Introduction}
The observed orbital period distribution of CVs shows a lack of systems in the 
$\sim\,2-3$\,h orbital period range, known as the period gap. 
In order to explain the gap, \citet{rappaportetal83-1} proposed the disrupted magnetic braking (DMB) scenario which today is frequently called the standard scenario for CV evolution. The main assumption of the DMB scenario is that MB turns off when the donor star becomes fully convective at $\Porb\sim$\,3h. 
Systems above the gap evolve towards shorter orbital periods, 
due to gravitational radiation (GR) and efficient MB. Due to the strong mass transfer caused by MB the donors are driven out of thermal equilibrium. Once the donor star becomes fully convective, at the upper edge of the gap, MB stops or at least becomes inefficient, causing a drop in the mass-transfer rate, which allows the donor star to relax to a radius which is smaller than its Roche lobe radius. The system detaches, mass transfer stops and the system becomes a detached white dwarf plus main sequence (WD+MS) binary, evolving towards shorter periods only via GR. At orbital periods of $\sim$\,2h the system is close enough to restart mass transfer and the system appears again as a CV at the lower boundary of the gap.
It has been shown frequently that DMB can well explain the orbital period gap
\citep[e.g.][]{howelletal01-1}. Independent evidence for a dramatic change of the strength of MB comes from observations of single stars, as well as WD+MS binaries \citep{schreiberetal10-1,rebassa-mansergasetal13-1} 

However, theoretical binary population synthesis (BPS) models of CVs based on DMB also reveal dramatic disagreements between observations and theoretical predictions:
1.\, CVs below the period gap should make up $\sim$\,99 per cent of the population
\citep{kolb93-1}, while roughly the same number of systems is observed 
below and above the gap \citep{knigge06-1}.  
2.\, The observed orbital period minimum is at 
76\,min \citep{knigge06-1}, 
while the predicted orbital period minimum 
is 65-70\,min \citep{kolb+baraffe99-1}. 
3.\, A strong accumulation of systems at the orbital 
period minimum is predicted, but not observed 
\citep{patterson98-1,knigge06-1}. 
4.\, The predicted space density \citep[e.g.][]{ritter+burkert86-1,kolb93-1} exceeds the observed one by 1-2 orders of magnitude
\citep[e.g.][]{patterson98-1,pretorius+knigge12-1}. 
5.\, BPS models predict an average white dwarf (WD) 
mass of $\lesssim0.6\Msun$
\citep[e.g.][]{politano96-1} and large numbers of 
CVs containing helium core (He-core) WDs. 
Yet, the observed average
mass is $\sim0.83\Msun$ and not a single system 
with a definite He-core WD primary 
has been identified so far.
The WD mass problem is probably the most severe 
as observational biases can be clearly excluded to play a significant role 
\citep{zorotovicetal11-1}. 

During the past decade, one of these long standing 
problems of CV evolution has been solved.  
The problem of the missing spike at the 
orbital period minimum disappeared when 
recent deep surveys identified a previously 
hidden population of faint CVs \citep{gaensickeetal09-2}. 
Furthermore, significant progress has been made with the orbital period minimum problem. 
If additional angular 
momentum loss, apart from GR, is 
assumed to be present in systems with fully 
convective donor stars, the predicted period minimum is 
shifted to longer orbital periods as it is observed 
\citep[e.g.][]{kniggeetal11-1}.  
However, the cause of the additional braking mechanisms remains unclear. 

Here we present BPS models of CVs using the standard 
model for CV evolution but taking into account additional angular momentum loss 
generated by the mass transfer in CVs.  
This consequential angular momentum 
loss (CAML) can drive CVs with low-mass WDs
into dynamically unstable mass transfer
which solves the WD mass problem, the space density problem, and 
also brings into agreement the predicted and observed 
orbital period distributions. Finally, the proposed CAML might
be the same missing angular momentum loss mechanism that is 
required to shift the theoretically predicted orbital period minimum to 
longer periods. The idea of CAML potentially solving simultaneously several 
major problems of CV evolution has been first presented 
at the CV conference in New York (November 2014) by the lead author 
of this article and is published in 
\citet{schreiberetal16-1}. A nice complementary paper partly inspired by the 
New York conference is in press \citep{nelemansetal15-1}.

%
%

\section{The WD mass problem}
The currently most severe problem with CV evolution is the white dwarf mass problem. Four possible solutions for the WD mass problem have been discussed in the literature: 
\begin{itemize}
\item The observed WD mass distribution is the result of an observational bias. This idea goes back to \citet{ritter+burkert86-1} who were able to explain the early measurements of large WD masses in the bright CVs discovered decades ago. However, a significant fraction of CVs more recently identified by SDSS are dominated by emission from the WD and the observed sample of such systems should be biased towards low mass WDs 
while the measured masses still cluster around $0.8\Msun$ \citep{zorotovicetal11-1}.
\item The low mass WDs in CVs accrete more mass than is expelled during the subsequent nova eruption. As shown by \citet{wijnenetal15-1}, mass growth can not solve the problem of the large WD masses in CVs because most of the predicted CVs with low mass WDs have also low mass companions. Even assuming that the total mass of the secondary would be accreted by the WD, which is of course nonsense, does not allow to reproduce the observed distribution with a mean mass of $\sim0.8\Msun$ \citep{zorotovicetal11-1}.  
\item Systems that would form CVs with low mass WDs do not survive CE evolution. 
This idea has been put forward by \citet{geetal15-1}. However, as shown by \citet{zorotovicetal11-1}, the existence of detached post common envelope binaries with low mass WDs clearly excludes this option. 
\item The WDs in CVs grow in mass before they become CVs. This option 
has been investigated by \citet{wijnenetal15-1}. Indeed, if a large number of CVs would form from thermal time scale mass transfer, the predicted WD mass distribution would become similar to the observed one. However, this approach would predict even more CVs than the old models with disastrous consequences for the space density problem. In addition, a large number of CVs descending from thermal time scale mass transfer would predict a large number of CVs with evolved secondary stars which contradicts the observations
\citep{gaensickeetal03-1}. 

\end{itemize}
All the above approaches fail to explain the observed WD mass distribution or 
generate dramatic disagreement with other features of the observed CV sample. Here we show that one frequently overlooked but crucial and uncertain assumption in binary population models might be responsible for the disagreement between theory and observation. The relatively low mass transfer rates of CVs 
imply that mass transfer is driven by angular momentum loss. In other words, CVs must be stable against thermal and dynamical time scale mass transfer.  
The stability limit for mass transfer is therefore a key ingredient of binary population models. It depends crucially on the mass ratio of the system and the mass and angular momentum that is lost as a consequence of the mass transfer.
In the following section we show that the WD mass problem and several other problems can be solved if CAML drives CVs with low mass WDs into unstable mass transfer.

\section{Binary population models}
\begin{figure*}
\includegraphics[width=1.0\textwidth]{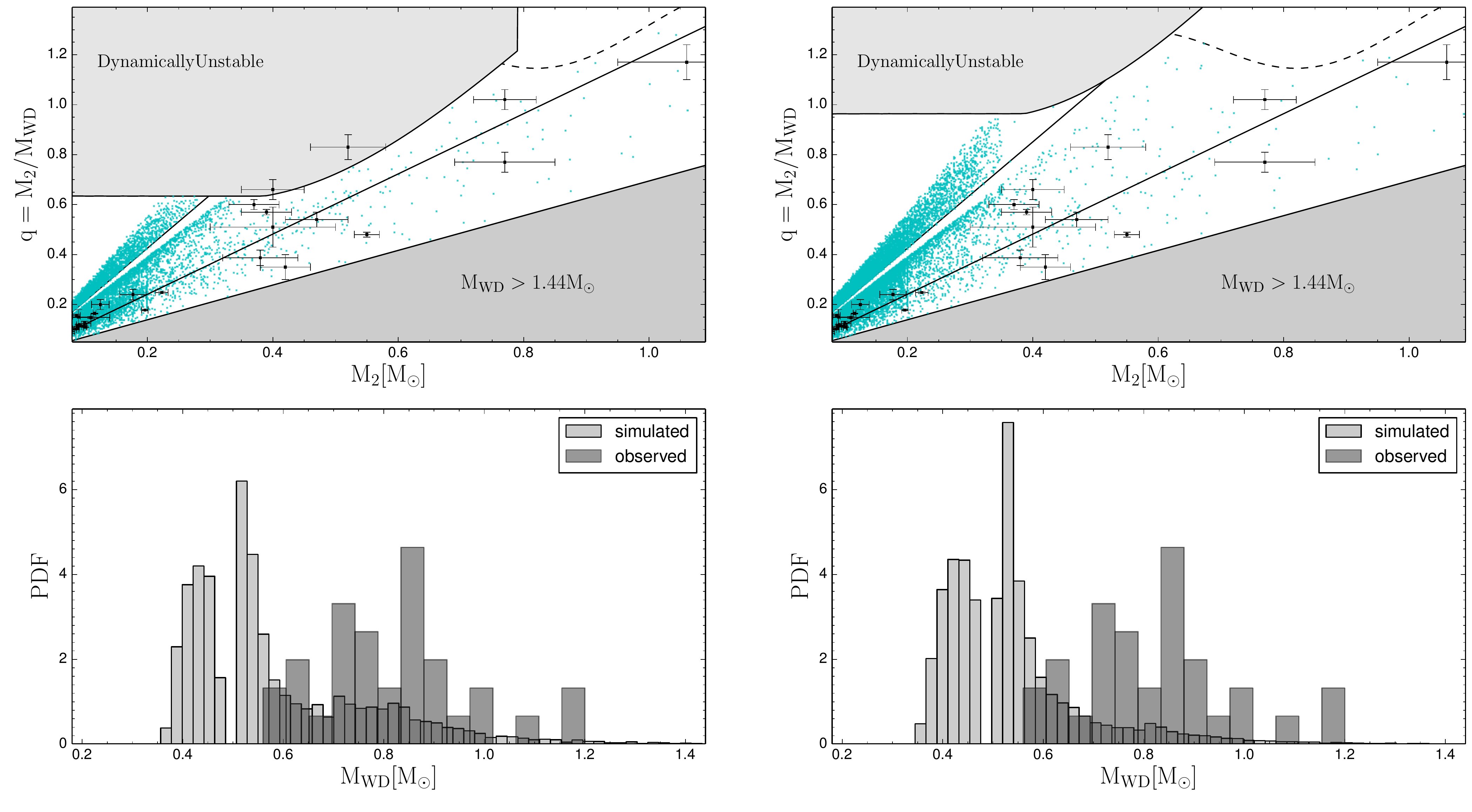}
\caption{\textit{Top}: Observed 
(black squares) and predicted (cyan dots)
CV populations in the q versus $M_2$ diagram 
for the fully conservative case (\textit{left}) 
and for the classical non-conservative model 
(\textit{right}). 
The grey shaded areas represent the forbidden regions either due to 
dynamically unstable mass
transfer or because the WD mass exceeds the Chandrasekhar limit. 
The black solid lines represent the average of the WD masses measured in CVs 
($\Mwd=0.83\Msun$) and the mass limit for He-core WDs ($0.47\Msun$). 
The dashed line represents the limit for thermally unstable mass transfer. 
\textit{Bottom}: Comparison between the observed (dark grey) and simulated
(light grey) WD mass distribution. Apparently, both models can not reproduce the observations. 
}
\label{fold}
\end{figure*}
To illustrate the impact of the stability limit we performed BPS calculations
with different assumptions for mass and angular momentum loss. Details concerning our model assumptions (concerning e.g. the initial mass ratio distribution, common envelope evolution, and the initial mass function) can be found in \citet{schreiberetal16-1}. 

The stability limit can be calculated by 
equating the adiabatic mass radius exponent of the secondary star and the mass 
radius exponent of its Roche-lobe.   
The first depends only on the 
structure of the secondary star and has been calculated by \citet{hjellming89-1}. 
The second, however, depends on the change of the mass ratio and 
the amount of angular momentum loss generated by mass transfer. 
In previous binary population models of CVs either the fully conservative case (the total mass and angular momentum of the binary is conserved) or a weak form of CAML was used. In the latter case, it is assumed that during nova eruptions exactly the amount of mass that has been previously accreted by the WD is expelled and carries the specific angular momentum of the WD. We performed binary population models for both these classical assumptions in the framework of the DMB scenario and the results are shown in Fig.\,1. Clearly, both models dramatically fail to reproduce the WD mass distribution. 
However, Fig.\,1 also illustrates how strong the stability limit depends on the assumptions concerning mass and angular momentum 
loss generated by mass transfer. In the fully conservative case less CVs survive because 
the mass of the white dwarf is assumed to grow which reduces the Roche-volume of the secondary. This effect dominates over the reduced angular momentum loss which makes the secondaries Roche-lobe larger (but to a lesser extent). 

Given that the stability limit sensitively depends on the assumptions 
of CAML and mass loss we investigate if and which form of CAML could change the stability limit in a way that solves the WD mass problem. We assume mass loss during nova eruption 
keeps the WD mass constant during secular evolution (exactly the same amount of mass that has been previously accreted is expelled in a nova event). However, in contrast to previous works we keep the angular momentum loss associated with mass loss as a free parameter. 
Given that the disagreement between 
the observed and predicted WD masses in CVs 
is mainly caused by the large number of predicted
CVs with low-mass WDs with low mass companions
(see Fig.1), we expect significant 
improvement if consequential angular momentum loss is stronger 
for low mass WDs. 

\begin{figure*}
\includegraphics[width=1.0\textwidth]{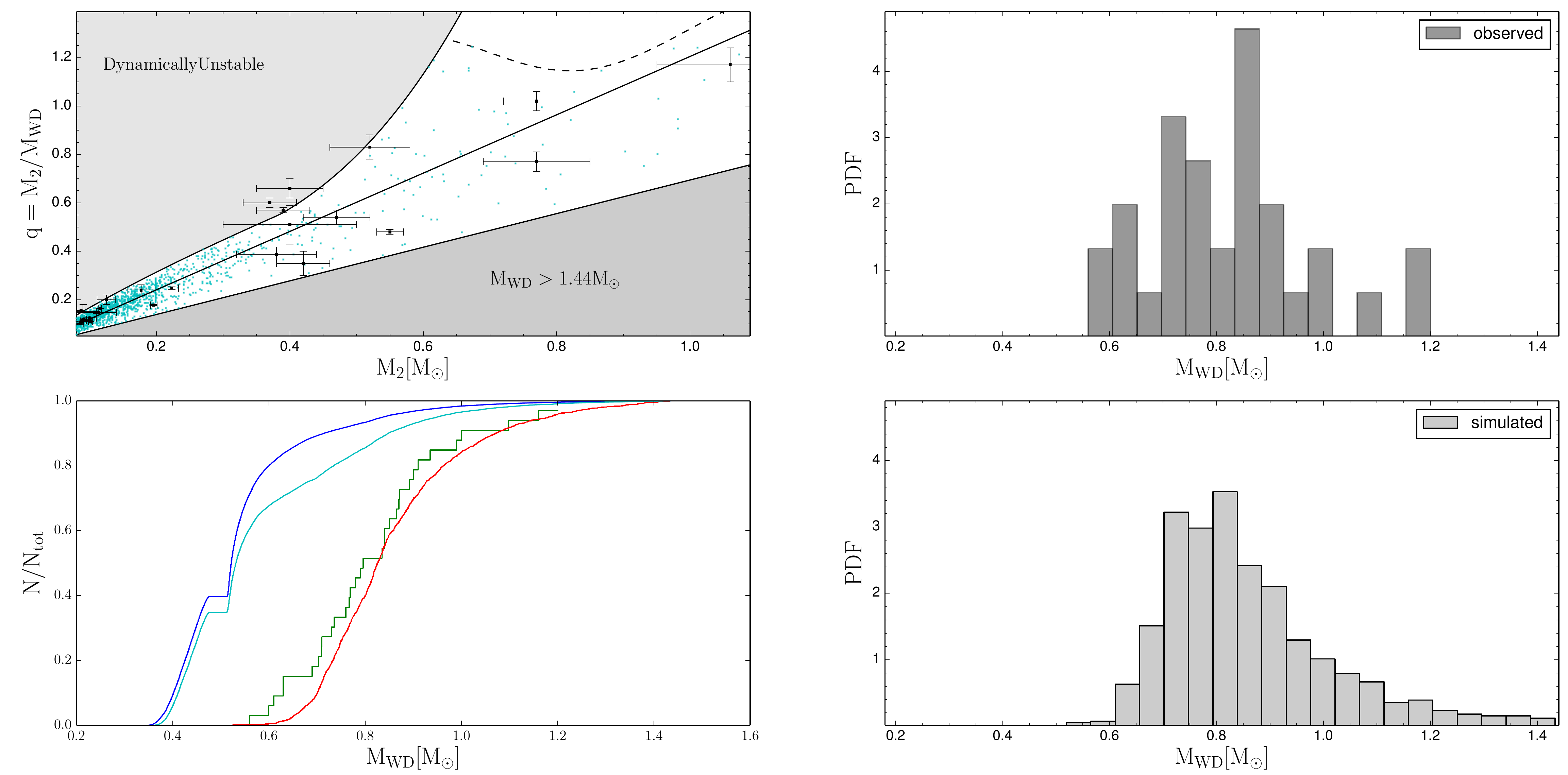}
\caption{\textit{Top left}: Observed and predicted
CV populations in the q versus $M_2$ diagram 
for our empirical CAML model. Colours and symbols
are the same as in Fig.\,1.
\textit{Right}: Observed (\textit{top}) and simulated 
(\textit{bottom}) WD mass distribution for the empirical 
CAML model.
\textit{Bottom left}: Cumulative distribution
of WD masses in CVs for the observed systems (green)
and for our three models: 
fully conservative (cyan), 
classical non-conservative (blue) and empirical 
CAML (red). Clearly, the latter model can 
reproduce 
the observations while the predictions of the other two models 
dramatically disagree with the observations.}
\label{fnew}
\end{figure*}

Indeed, we find good agreement if we assume
decreasing CAML for increasing WD masses 
\citep[for details see][]{schreiberetal16-1}. 
In the top left panel of 
Fig.\,2 we show 
the observed and predicted CV populations in the $q$ 
versus $M_2$ diagram for a simulated population of 
CVs assuming this empirical CAML (eCAML from now on)
model. 
Apparently, eCAML can reproduce the observed 
WD mass distribution in CVs very well. 
However, before the eCAML model can be considered 
a viable option for CV evolution, 
we need to investigate how its predictions compare 
with other observed properties of CVs and we need to 
find a physical mechanism that might be 
responsible for eCAML. 
  
\section{Orbital period distribution 
and space density}
\begin{figure}
\begin{center}
\includegraphics[angle=270, width=0.8\textwidth]{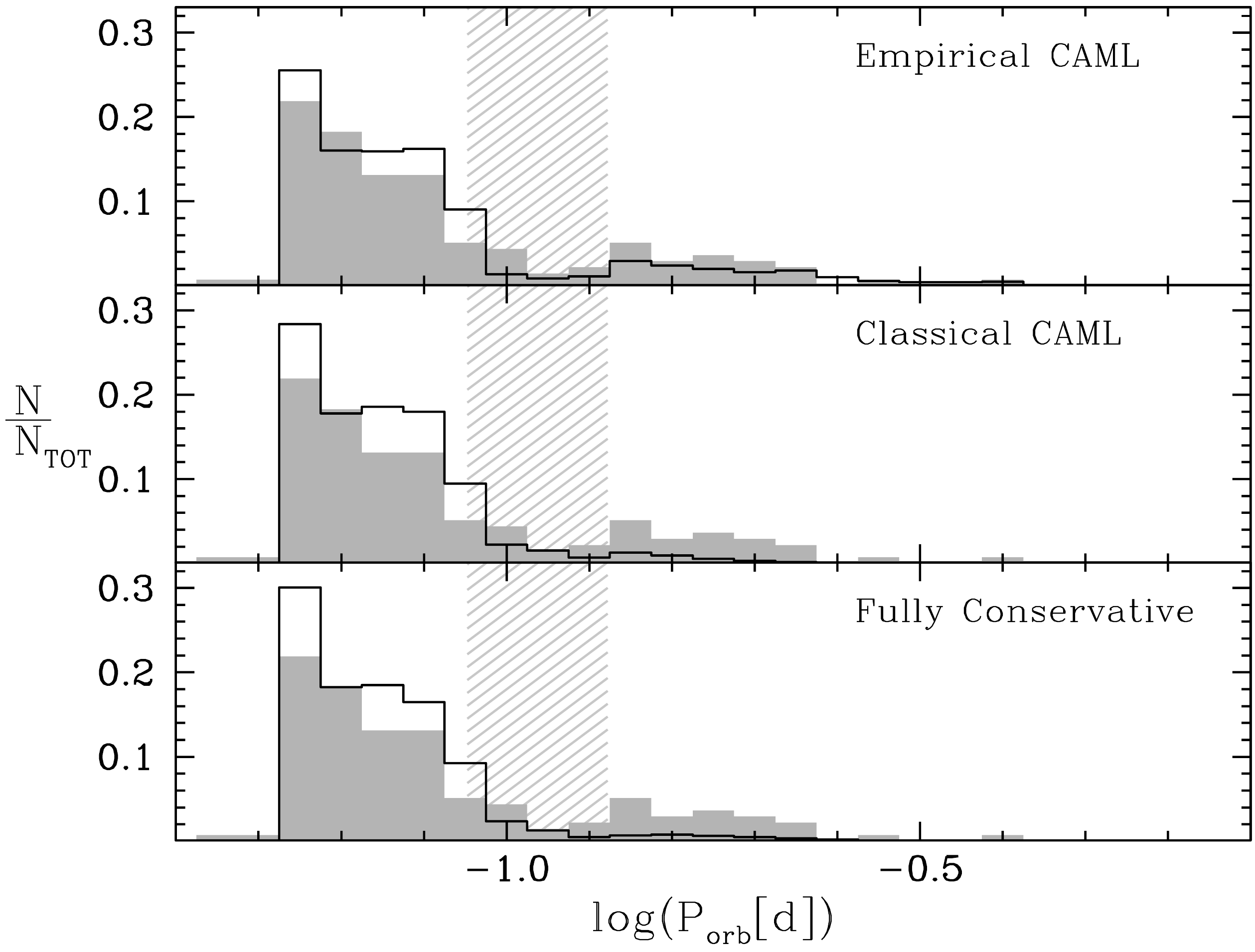}
\caption{Orbital-period distributions for our three 
models of CV population: 
eCAML (\textit{top}), classical CAML (\textit{middle}), 
and the fully conservative (\textit{bottom}). In grey the observed orbital
period as measured by \citet{gaensickeetal09-2}. The eCAML model is not only
the first model able to reproduce the observed WD mass distribution 
but also the first one to be in agreement with the 
observed orbital period distribution. }
\label{fporb}
\end{center}
\end{figure}
As described in the introduction, previous 
binary population models of CVs failed 
to reproduce both the predicted space density of CVs 
and the orbital period distribution. 
We now show that incorporating eCAML not only 
solves the WD mass problem but also brings into 
agreement the predicted and observed space density 
and orbital period distribution of CVs. 

In Fig.\,3 we compare 
the orbital period distribution of SDSS CVs 
\citep{gaensickeetal09-2}
with those predicted by the three models 
discussed in this paper. 
The two old models do not only dramatically disagree 
with the WD mass distribution but also 
predict the existence of too many CVs 
at short orbital periods and too few
above the gap. In contrast, the eCAML model that was 
designed to solve the WD mass problem agrees 
very well with the observed period distribution.

The period distribution problem is not the only 
long standing issue the eCAML approach
solves simultaneously with the WD mass problem. 
As a large number of systems containing 
low-mass WDs is supposed to suffer from unstable 
mass transfer, the total number of CVs 
predicted by the eCAML model is about one order of magnitude 
smaller than in previous models which brings the predicted CV space density 
into the range of the values derived from observations.

\section{Physical interpretation of eCAML}
The eCAML model presented in this paper can 
simultaneously solve the three biggest problems 
of CV evolution. However, there must be 
a physical mechanism behind this parameterised 
model. 
The most obvious mechanism that may generate 
significant CAML in CVs are nova eruptions. 
In fact, the classical CAML model represents
a weak form of CAML caused by novae.
This model assumes the 
expelled material to carry away the specific 
angular momentum of the WD which represents a reasonable assumption
\emph{if} friction between the ejecta and the 
secondary star is negligible. If, 
on the other hand, friction contributes
significantly, CAML might be much stronger 
than predicted by the classical CAML model. 
In other words, when a CV formed from a detached post common envelope binary, the first nova may lead to a common envelope like situation in which a significant amount of orbital energy and angular momentum is used to expel the nova shell which causes the two stars to merge \citep[see also][]{nelemansetal15-1}.  

As shown by \citet{schenkeretal98-1}, 
frictional angular momentum loss produced 
by novae only weakly depends on the 
mass ratio but is very sensitive to the 
expansion velocity of the ejecta, 
i.e. assuming slow expansion velocities,
the consequential (frictional) angular 
momentum loss can be significant
\citep[see][their Figure 5]{schenkeretal98-1}.  
Interestingly, nova models predict significantly 
faster expansion velocities for CVs containing 
high-mass WDs than for those containing low-mass 
WDs \citep[see e.g.][their Fig.\,2c]{yaronetal05-1}. 
The CAML generated by slow nova eruptions in CVs 
containing low-mass WDs might therefore 
be the physical cause behind the 
eCAML model we used to bring into 
agreement observations and theory.

\section{Conclusion} 
For decades we used simple assumptions like the full conversation of mass and angular momentum to determine the stability limits of mass transfer in CVs. 
While these assumptions have been reasonable when no additional information was available, with thousands of CVs known and with accurate stellar and binary parameters measured for many of them, now the time has come to use observational constraints to improve our models of compact binary evolution. 
We used measurements of the white dwarf masses 
in CVs to define an empirical description of consequential angular momentum loss that brings into agreement observation and theory. The key prediction of the new scenario is that post common envelope binaries containing low mass white dwarfs merge quickly after the secondary filled its Roche-lobe. This new model for CV evolution, first presented by the lead author of this paper at the CV conference in New York  (November, 2014) and described in detail in \citet{schreiberetal16-1}, simultaneously solves several major problems of CV evolution. It is the only model presented so far that reproduces the observed white dwarf mass distribution. In addition, 
the predicted space density and the relative number of systems below the gap are significantly reduced which solves the space density and the orbital period distribution problem. Finally, the additional angular momentum loss 
may shift the predicted orbital period minimum to longer orbital periods
solving the orbital period minimum problem \citep[see also][]{nelemansetal15-1}.

\bibliographystyle{mn2e}

\begin{thebibliography}{22}
\expandafter\ifx\csname natexlab\endcsname\relax\def\natexlab#1{#1}\fi

\bibitem[{{G{\" a}nsicke} {et~al}\mbox{.}(2003){G{\" a}nsicke}, {Szkody}, {de
  Martino}, {Beuermann}, {Long}, {Sion}, {Knigge}, {Marsh}, \&
  {Hubeny}}]{gaensickeetal03-1}
{G{\" a}nsicke} B.~T. {et~al.}, 2003, \apj, 594, 443

\bibitem[{{G{\"a}nsicke} {et~al}\mbox{.}(2009){G{\"a}nsicke}, {Dillon},
  {Southworth}, {Thorstensen}, {Rodr{\'{\i}}guez-Gil}, {Aungwerojwit}, {Marsh},
  {Szkody}, {Barros}, {Casares}, {de Martino}, {Groot}, {Hakala}, {Kolb},
  {Littlefair}, {Mart{\'{\i}}nez-Pais}, {Nelemans}, \&
  {Schreiber}}]{gaensickeetal09-2}
{G{\"a}nsicke} B.~T. {et~al.}, 2009, \mnras, 397, 2170

\bibitem[{{Ge} {et~al}\mbox{.}(2015){Ge}, {Webbink}, {Chen}, \&
  {Han}}]{geetal15-1}
{Ge} H., {Webbink} R.~F., {Chen} X., {Han} Z., 2015, \apj, 812, 40

\bibitem[{{Hjellming}(1989)}]{hjellming89-1}
{Hjellming} M.~S., 1989, PhD thesis, AA(Illinois Univ.~at Urbana-Champaign,
  Savoy.)

\bibitem[{{Howell}, {Nelson} \& {Rappaport}(2001){Howell}, {Nelson}, \&
  {Rappaport}}]{howelletal01-1}
{Howell} S.~B., {Nelson} L.~A., {Rappaport} S., 2001, \apj, 550, 897

\bibitem[{{Knigge}(2006)}]{knigge06-1}
{Knigge} C., 2006, \mnras, 373, 484

\bibitem[{{Knigge}, {Baraffe} \& {Patterson}(2011){Knigge}, {Baraffe}, \&
  {Patterson}}]{kniggeetal11-1}
{Knigge} C., {Baraffe} I., {Patterson} J., 2011, \apjs, 194, 28

\bibitem[{{Kolb}(1993)}]{kolb93-1}
{Kolb} U., 1993, \aap, 271, 149

\bibitem[{{Kolb} \& {Baraffe}(1999)}]{kolb+baraffe99-1}
{Kolb} U., {Baraffe} I., 1999, \mnras, 309, 1034

\bibitem[{{Nelemans} {et~al}\mbox{.}(2015){Nelemans}, {Siess}, {Repetto},
  {Toonen}, \& {Phinney}}]{nelemansetal15-1}
{Nelemans} G., {Siess} L., {Repetto} S., {Toonen} S., {Phinney} E.~S., 2015,
  ArXiv e-prints:1511.07701

\bibitem[{{Patterson}(1998)}]{patterson98-1}
{Patterson} J., 1998, PASP, 110, 1132

\bibitem[{{Politano}(1996)}]{politano96-1}
{Politano} M., 1996, \apj, 465, 338

\bibitem[{{Pretorius} \& {Knigge}(2012)}]{pretorius+knigge12-1}
{Pretorius} M.~L., {Knigge} C., 2012, \mnras, 419, 1442

\bibitem[{{Rappaport}, {Joss} \& {Verbunt}(1983){Rappaport}, {Joss}, \&
  {Verbunt}}]{rappaportetal83-1}
{Rappaport} S., {Joss} P.~C., {Verbunt} F., 1983, \apj, 275, 713

\bibitem[{{Rebassa-Mansergas}, {Schreiber} \&
  {G{\"a}nsicke}(2013){Rebassa-Mansergas}, {Schreiber}, \&
  {G{\"a}nsicke}}]{rebassa-mansergasetal13-1}
{Rebassa-Mansergas} A., {Schreiber} M.~R., {G{\"a}nsicke} B.~T., 2013, \mnras,
  429, 3570

\bibitem[{{Ritter} \& {Burkert}(1986)}]{ritter+burkert86-1}
{Ritter} H., {Burkert} A., 1986, \aap, 158, 161

\bibitem[{{Schenker}, {Kolb} \& {Ritter}(1998){Schenker}, {Kolb}, \&
  {Ritter}}]{schenkeretal98-1}
{Schenker} K., {Kolb} U., {Ritter} H., 1998, \mnras, 297, 633

\bibitem[{{Schreiber} {et~al}\mbox{.}(2010){Schreiber}, {G{\"a}nsicke},
  {Rebassa-Mansergas}, {Nebot Gomez-Moran}, {Southworth}, {Schwope},
  {M{\"u}ller}, {Papadaki}, {Pyrzas}, {Rabitz}, {Rodr{\'{\i}}guez-Gil},
  {Schmidtobreick}, {Schwarz}, {Tappert}, {Toloza}, {Vogel}, \&
  {Zorotovic}}]{schreiberetal10-1}
{Schreiber} M.~R. {et~al.}, 2010, \aap, 513, L7

\bibitem[{{Schreiber}, {Zorotovic} \& {Wijnen}(2016){Schreiber}, {Zorotovic},
  \& {Wijnen}}]{schreiberetal16-1}
{Schreiber} M.~R., {Zorotovic} M., {Wijnen} T.~P.~G., 2016, \mnras, 455, L16

\bibitem[{{Wijnen}, {Zorotovic} \& {Schreiber}(2015){Wijnen}, {Zorotovic}, \&
  {Schreiber}}]{wijnenetal15-1}
{Wijnen} T.~P.~G., {Zorotovic} M., {Schreiber} M.~R., 2015, \aap, 577, A143

\bibitem[{{Yaron} {et~al}\mbox{.}(2005){Yaron}, {Prialnik}, {Shara}, \&
  {Kovetz}}]{yaronetal05-1}
{Yaron} O., {Prialnik} D., {Shara} M.~M., {Kovetz} A., 2005, \apj, 623, 398

\bibitem[{{Zorotovic}, {Schreiber} \& {G{\"a}nsicke}(2011){Zorotovic},
  {Schreiber}, \& {G{\"a}nsicke}}]{zorotovicetal11-1}
{Zorotovic} M., {Schreiber} M.~R., {G{\"a}nsicke} B.~T., 2011, \aap, 536, A42

\end{thebibliography}

%

\end{document}